\documentclass[12pt]{article}
\begin{document}
\begin{large}
\title{\bf {Typicality vs. probability in trajectory-based formulations of quantum mechanics}}
\end{large}
\author{Bruno Galvan \footnote{Electronic address: b.galvan@virgilio.it}\\ \small Loc. Melta 40, 38014 Trento, Italy.}
\date{\small January 2007}
\maketitle

\begin{abstract}
Bohmian mechanics represents the universe as a set of paths with a probability measure defined on it. The way in which a mathematical model of this kind can explain the observed phenomena of the universe is examined in general. It is shown that the explanation does not make use of the full probability measure, but rather of a suitable set function deriving from it, which defines relative typicality between single-time cylinder sets. Such a set function can also be derived directly from the standard quantum formalism, without the need of an underlying probability measure. The key concept for this derivation is the {\it quantum typicality rule}, which can be considered as a generalization of the Born rule. The result is a new formulation of quantum mechanics, in which particles follow definite trajectories, but which is based only on the standard formalism of quantum mechanics.
\end{abstract}

%begin{large}
\section{Introduction}
Bohmian mechanics is a complete and coherent formulation of non-relativistic quantum mechanics \cite{bohm1, bohm3, durr2, allori, durr1}. According to this formulation, the particles of the universe follow definite trajectories satisfying a differential equation, the guidance equation. The set of these trajectories is endowed with a probability measure deriving from the universal wave function. In spite of its completeness and coherence, Bohmian mechanics is far from a universally accepted formulation of quantum mechanics, the presence of unobservable entities like Bohmian trajectories being one of its most criticized features.

In short, we will refer to a generic set of paths with a probability measure defined on it as a {\it path space}. The way in which a path space can explain the observed phenomena of the universe is a very intriguing conceptual issue, and we think it has been only partially investigated in the literature. Most of the work in this sense has been done in the context of Bohmian mechanics \cite{durr1}.

A path space has the same structure as a stochastic process, and usually stochastic processes are utilized to represent ensembles of open systems, such as particles in a liquid subjected to Brownian motion. There are two fundamental differences when a path space represents the universe instead of an ensemble of open systems: (1) the observers are inside the system, and they cannot perform all the measurements allowed in the previous case; (2) there is just one universe, thus the probability measure is not used to derive relative frequencies, but rather typicality. 

As we will see, the consequences of these differences are that the full structure of the probability measure $\mu$ is unobservable, and that the explanation given by a path space is based on the set function 
\begin{equation}
r_\mu(S_1|S_2):=\frac{\mu(S_2 \setminus S_1)}{\mu(S_2)},
\end{equation}
where $S_1$ and $S_2$ are single-time cylinder sets. This set function will be referred to as the {\it relative typicality function}, because $r_\mu(S_1|S_2) \ll 1$ implies that $S_1$ is typical relative to $S_2$, i.e. the overwhelming majority of the paths of $S_2$ also belong to $S_1$.

The crucial point is that, while the quantum formalism cannot define a probability measure on a set of paths, in a natural way it can define a set function, the {\it mutual typicality function}, from which the relative typicality function can be derived. The mutual typicality function must be accompanied by an interpretative rule connecting it with typicality. This rule will be referred to as the {\it quantum typicality rule}, and it can be considered a generalization of the Born rule. The result is a new formulation of quantum mechanics, in which particles follows definite trajectories, as in Bohmian mechanics, but which is based only on the formalism of standard quantum mechanics, the guidance equation being replaced by the quantum typicality rule.

\vspace{3mm}
The paper is structures according to the following scheme. In section \ref{path} a formal definition, the main properties and some examples of path spaces are given. In section \ref{explanation} the way in which a path space explains the observed phenomena of the universe is studied, and it is shown that this explanation is based on the relative typicality function. In section \ref{quantum} the possibility to derive typicality functions from the quantum formalism is shown, and many related technical issues are discussed. In section \ref{discussion} there is a final discussion about the proposed formulation of quantum mechanics.

%newpage

\section{Path spaces} \label{path} 

In this section, the formal definition and the main properties of a space of paths with a probability measure defined on it are explained. Since such a structure is a stochastic process, most of the terminology and the properties of these spaces are derived from stochastic processes.

Let $(M, {\cal  B})$ be a measurable space, $T$ an index set and $\Lambda$ a set of mappings from $T$ to $M$. In this paper $T$ will always be the positive time axis $R^+$, and, with the exception of the example of the classical universe, $M$ will always be the configuration space $R^{3N}$ of an $N$-particle system. Given $t \in T$ and $\Delta \in {\cal  B}$, the subset $(t, \Delta):=\{\lambda \in \Lambda: \lambda(t) \in \Delta \}$ is a {\it single-time cylinder set} ({\it  s-set}, in short); a {\it cylinder set} is any finite intersection of s-sets. The shorthand notation $S_i$ will be used to denote the s-set $(t_i, \Delta_i)$, $i=1, 2, \ldots$. Let ${\cal  S}$ denote the class of the s-sets, and $\sigma({\cal  S})$ the $\sigma$-algebra generated by ${\cal  S}$. 

A {\it path space} is the pair $(\Lambda, \mu)$, where $\mu$ is a probability measure on $\sigma({\cal  S})$. A path space is defined to be {\it canonical} if $\Lambda=M^T$, where $M^T$ is the set of all the mappings from $T$ to $M$.

By defining $z_t(\lambda):=\lambda(t)$, $\{z_t\}_{t \in T}$ is then a class of random variables on the probability space $(\Lambda, \sigma({\cal  S}), \mu)$, indexed by $T$. Thus any path space $(\Lambda, \mu)$ naturally corresponds to the stochastic process $(\Lambda, \sigma({\cal  S}), \mu, \{z_t\}_{t \in T})$\footnote{The converse is not true in general: given a stochastic process $(\Omega, {\cal F}, \mu, \{z_t\}_{t \in T})$, every element $\omega \in \Omega$ defines the sample path $\lambda(t):=z_t(\omega)$, but the correspondence between $\Omega$ and the set of the sample paths may be non-biunivocal. However, by definition, a canonical stochastic process is also a path space.}.

The values of the measure $\mu$ on the cylinder sets are the {\it finite dimensional distributions } of the path space, while its value on the s-sets is the {\it single-time distribution}. Two path spaces with the same index set and state space are said to be {\it equivalent} if they have the same finite dimensional distributions. Two path spaces with the same set $\Lambda$ and the same finite dimensional distributions are identical, i.e. they have the same probability measure $\mu$. Any class of finite dimensional distributions satisfies some formal relations. According to the Kolmogolov reconstruction theorem, given any class of finite dimensional distributions satisfying such relations, there exists a unique canonical path space giving rise to that class of finite dimensional distributions. 

We say that a path space is {\it deterministic} if for any s-set $(t_1, \Delta_1)$ and any $t_2 \in T$ there exists $\Delta_2 \in {\cal B}$ such that $\mu[(t_1, \Delta_1) \triangle (t_2, \Delta_2)]=0$, where $ \triangle $ is the symmetric difference.

\vspace{3mm} We now give some examples of path spaces. \\ {\it Classical system.} The state space $M$ is the phase space of a classical Hamiltonian system. Let $\Delta$ be a subset of $M$ with $0 < \mu_L(\Delta) < \infty$, where $\mu_L$ is the Lebesgue measure on $M$. The set $\Lambda$ is composed of the Hamiltonian trajectories $\lambda: R^+ \rightarrow M$ such that $\lambda(0) \in \Delta$. The measure $\mu_C$ on $\Lambda$ is defined by
\begin{equation}
\mu_C(\Gamma):=\frac{\mu_L[z_t(\Gamma)]}{\mu_L(\Delta)}, \; \; \; \Gamma \in \sigma({\cal S}).
\end{equation}
Due to the Liouville theorem, the above definition does not depend on the time. This path space is deterministic.

\vspace{3mm}

{\it Bohmian mechanics}. Hereafter the state space $M$ will always be the configuration space $R^{3N}$ of an N-particle system. Let us assume a normalized universal wave function $\Psi(t)=U(t)\Psi_0 \in L^2(M)$, where $U(t)$ is the unitary time evolution operator. $\Lambda$ is the set of the trajectories satisfying the guidance equation 
\begin{equation}
\frac{d{\bf x}_k}{dt}=\frac{\hbar}{m_k}\hbox{Im}\frac{{\bf \nabla}_k \Psi}{\Psi}, \; \; k=1, \ldots, N.
\end{equation}
The measure $\mu_B$ is defined by
\begin{equation} \label{bohmmeasure}
\mu_B(\Gamma):=||E[z_t(\Gamma)]\Psi(t)||^2,
\end{equation}
where $E(\Delta)$ is the spatial projector onto $\Delta \in {\cal B}$. Due to the equivariance property of Bohmian mechanics, the above definition does not depend on the time. The single-time distribution of Bohmian mechanics is 
\begin{equation} \label{md}
\mu_B[(t, \Delta)] = ||E(\Delta) \Psi(t)||^2.
\end{equation}
Bohmian mechanics is deterministic.

\vspace{3mm}
{\it The Everett-Bell universe.}

The set $\Lambda$ is $M^T$, and the measure $\mu_E$ is defined by the finite dimensional distributions
\begin{equation} \label{simsp}
\mu_E(S_1\cap  \ldots \cap S_n)
 :=  ||E(\Delta_1) \Psi(t_1)||^2 \ldots ||E(\Delta_n) \Psi(t_n)||^2,
\end{equation}
where the assumption is made that $t_i \neq t_j$ for $i \neq j$. This universe was introduced, although in a less formal way, by Bell \cite{bell1, bell2}, as a version of the relative state formulation of quantum mechanics by Everett. This universe is very unphysical, because it has no dynamics, i.e. no law connecting configurations at different times, and it defines no physical trajectory.

\vspace{3mm}
{\it The impossible quantum path space}. One could try to define the following ``quantum'' finite dimensional distributions:
\begin{equation} \label{notadd} 
\mu_Q (S_1\cap  \ldots \cap S_n):=|| E(\Delta_n) U(t_n-t_{n-1})E(\Delta_{n-1}) \ldots E(\Delta_1) U(t_1) \Psi_0||^2, 
\end{equation}
where the assumption is made that $t_1 \leq t_2 \leq \ldots \leq t_n$. This definition derives from the Born rule and from the reduction postulate, according to which it corresponds to the quantum mechanical probability to find the trajectory in the regions $\Delta_i$ at the times $t_i$, for $i=1, \ldots, n$. The problem is that this expression is not additive, i.e.
\begin{eqnarray*}
& & \mu_Q(S_1 \cap \ldots \cap (t_i, \Delta_i \cup \Delta'_i) \cap \ldots \cap S_n) \ne \\
& & \mu_Q (S_1 \cap \ldots \cap (t_i, \Delta_i)\cap  \ldots \cap S_n) + \mu_Q (S_1 \cap \ldots \cap (t_i, \Delta'_i)\cap  \ldots \cap S_n),
\end{eqnarray*}
and thus it cannot be a consistent class of finite dimensional distributions. This is the paradoxical aspect of the superposition principle of quantum mechanics, which prevents an open quantum systems from being represented by a path space (or by a stochastic process). As we will see, the situation changes when the system is the universe.

%newpage
\section{Paths spaces and explanation} \label{explanation} 

In this section we study the way in which a path space representing the universe can explain the observed phenomena. As an example, let us consider first the case in which the path space represents an ensemble of open systems, such as particles in a liquid subjected to Brownian motion. In this case, a natural assumption is that all and only the possible measurements which can be performed on the systems are finite sequences of position measurements at different times, in such a way that any one of these measurements corresponds to a cylinder set. Assume that the experimenter performs the same measurement on all the systems of the ensemble. Then the path space {\it explains} the relative frequency of the ``yes'' outcomes of the measurements if such a frequency is approximately equal to the probability measure of the cylinder set corresponding to the measurements. Therefore, in this kind of explanation, the finite dimensional distributions are utilized, while the exact structure of the paths is not relevant.

This form of explanation is no longer valid when the path space represents the universe, for two reasons: (1) the universe is a closed system, and the observers are inside it; as we will see, the consequence is that the assumption that all and only the admissible measurements are position measurements at different times is no longer valid. (2) There is just one universe, therefore we cannot speak of relative frequency of the outcomes.

Thus a new form of explanation must be developed, and this will be the subject of this section. In this study, and in the rest of the paper, the universe will be considered as an idealized non-relativistic universe composed of N distinguishable spinless particles.

%newpage
\subsection{The Everett-Bell universe}

Our study starts with a discussion on the Everett-Bell universe. In spite of its very unphysical features, Bell claims that --at least from the formal point of view-- such a model of universe can explain the observed phenomena. The problem of course is the following: how can the Everett-Bell universe explain our perception of a definite past evolution if it does not define trajectories, i.e. if it does not provide any kind of correlation among the positions of the particles at different times? The Bell's answer is that ``we have no access to the past, but only to memories, and these memories are just part of the instantaneous configuration of the world'' \cite{bell1}. In other words, we have memories, i.e. information about the position of the particles in the past, only because such information are encoded in some way in the present configuration of our recording devices, possibly including the neurons of our brain. Thus our memories do not derive from the actual past evolution of the particles, but just from their present configuration. We think that such a position is very questionable, but it is useful for the time being to make this assumption, and to study the form of explanation deriving from it.

According to the above assumption, for every $x \in M$ it is possible to decide if this represents a correct configuration or not, i.e. if in $x$ are encoded the memories of a quasi-classical past evolution and the correct results for all the past statistical experiments. For example, let us suppose that a suitable configuration $x$ includes a laboratory in which a two-slit experiment has been performed, as resulting from the configuration of the laboratory which includes a video recording of the experiment; suppose moreover that, according to $x$, the image of the particles on the photographic plate of the screen does not correspond to the expected distribution with the interference fringes; than  $x$ is not a correct configuration.

Let $\Sigma \subseteq M$ denote the subset of $M$ composed of all the correct configurations. Then one can claim that the path space explains the observed phenomena of the universe if
\begin{equation} \label{ex1}
\mu[(t, \overline{\Sigma})] \leq \epsilon \ll 1, \; \forall t \in T,
\end{equation}
where the over-bar denotes the complement. In order to better justify such a claim, consider a generic set of times $\{t_1, \ldots, t_n \}$, and the $n$ corresponding random variables $X_i(\lambda):=\chi_\Sigma[\lambda(t_i)]$, where $\chi_\Sigma$ is the characteristic function of the set $\Sigma$. We have $E(X_i) \geq 1 - \epsilon$ and $\sigma^2(X_i) \leq \epsilon$. Consider also the random variable $Y_n:=\frac{1}{n}\sum_i X_i$. By using the formula for the variance of a sum and the Schwarz inequality, it is easy to check that
\begin{equation} \label{ex2}
E(Y_n) \geq 1 - \epsilon \; \; \hbox{and} \; \; \sigma^2(Y_n) \leq \epsilon.
\end{equation}
Since (\ref{ex2}) holds for any set $\{t_1, \ldots, t_n \}$, one can deduce that the overwhelming majority of the trajectories spend the overwhelming majority of the time inside the set $\Sigma$.

As to the explanation of statistical experiments, we can also consider the following reasoning. A statistical experiment consists of a long sequence of identical elementary experiments, such as the toss of a coin or the passage of a quantum particle through a screen with two slits. Let us consider a {\it specific} statistical experiment, i.e. an experiment performed in a specific place at a specific time; the experiment ends at the time $t$. The experimental setup must include a recording device which registers the outcomes of the elementary measurements (in the two slit experiment this device is simply a photographic plate behind the screen). Let $\Delta \subset M$ be the set of the configurations representing a universe at the time $t$ in which that experiment has been performed. The set $\Delta$ includes the configurations corresponding to all the possible results for the elementary measurements. For instance, in the case of the coin tosses it also includes the sequence with all heads, and in the case of the two slit experiment it includes all possible distributions of the particles on the photographic plate. Let $\Delta' \subset \Delta$ be the set of the configurations corresponding to the correct results, i.e. an equal distribution for heads and tails in the coin toss experiment, and the interference fringes in the two slit experiment. Again we can claim that the path space explains these results if 
\begin{equation} \label{ex3}
\frac{\mu[(\Delta \setminus \Delta', t)]}{\mu[(\Delta, t)]} \ll 1.
\end{equation}
The explanations expressed by conditions (\ref{ex1}) and (\ref{ex3}) are based on the fact that the correct results are {\it typical}, i.e. they are the overwhelming majority of the possible results. See for instance \cite{goldstein} for a discussion of the validity of such a explanation. This explanation is analogous to the one given for the second law of thermodynamics \cite{goldstein} and for the quantum equilibrium hypothesis \cite{durr1}.

\vspace{3mm}
One can see that only a small part of the structure of a path space is involved in this explanation. Namely, what is needed is just the information that two sets $\Delta$ and $\Delta'$ satisfy (\ref{ex3}) (note that (\ref{ex1}) represent a particular case of (\ref{ex3})). This means that only the single-time distribution of the path space is relevant, being however redundant, while the correlations at different times given by the finite dimensional distributions are totally irrelevant. The structure of the paths is of course also irrelevant.

This is the reason why, according to this approach, the Bell-Everett universe has enough structure to explain the observed phenomena. Bell, after explaining why this model of the universe can work, claims that it cannot be taken seriously \cite{bell2}. His opposition is however on the philosophical level rather than on the logical one; at the same time, we argue that there is also a logical reason to reject it. The reason is that it is impossible for a universe without any dynamics to allow us to have memories, because the memories encoded in the present configuration cannot be decoded without making use of a dynamical law, i.e. of a law correlating configurations at different times. For instance, if we have a film reproducing the fall of a stone, in order to extract the true trajectory of the stone from the film we must make a number of dynamical assumptions: we must assume that light rays travel along straight lines, we must know the laws of refraction to understand the behaviour of light inside the lens of the camera, and so on. If the Bell approach were correct, dynamics could be deduced from only a knowledge of the set $\Sigma$; for instance, Newton's second law could be extracted from the Everett-Bell universe. Bell does not provide any method to do this, and we argue that such a method does not exist. On the contrary, we propose that a dynamics does exist, and that memories depend on it. The study of memories and of their dependence on dynamics will be the subject of the next subsection.
%newpage
\subsection{Memory and knowledge} \label{sec:memory}

A very natural requirement for memories is that they correspond to  what actually happened (of course this is not the case for the Everett-Bell universe). In order to express such a requirement in a mathematical form, let us suppose that a subset $\Delta_2 \subseteq M$ represents the knowledge that an observer has about the configuration of the universe at a time $t_2$. Than the observer can remember that at a time $t_1 < t_2$ the configuration of the universe was in a suitable set $\Delta_1 \subseteq M$ only if 
\begin{equation} \label{hi1}
\frac{\mu(S_2 \setminus S_1)}{\mu(S_2)} \ll 1,
\end{equation}
where, as usual, $S_i=(t_i, \Delta_i)$. Indeed, suppose that (\ref{hi1}) does not hold. This means that a non-infinitesimal part of the trajectories of $S_2$ does not come from $S_1$. Therefore the observer at the time $t_2$ cannot remember that at the time $t_1$ the configuration of the universe was in $\Delta_1$ because there is a non-negligible probability that this fact never happened. Thus condition (\ref{hi1}) corresponds to the requirement that {\it only what (almost) surely happened can be remembered}.

An immediate consequence of this reasoning is that observers cannot ``measure'' generic cylinder sets. For instance, an observer can measure the cylinder set $S_1 \cap S_2$, with $t_1 < t_2$, only if (\ref{hi1}) is satisfied. Indeed such a measurement requires that the observer knows that at the time $t_2$ the configuration of the universe belongs to $\Delta_2$, and that he remembers that at the time $t_1$ the configuration belonged to $\Delta_1$; this requirement implies the condition (\ref{hi1}). This conclusion is very important, and it implies that most of the structure of the probability measure $\mu$ is unobservable.

\vspace{3mm}
As to knowledge, a natural requirement is that {\it what can be known is only what can be remembered for a suitable amount of time}. We are thus led to the notion of branch. Let us represent the knowledge, evolving with time, that an observer has about the configuration of the universe as a mapping $h: I \rightarrow {\cal B}$, where $I$ is a time interval, and ${\cal B}$ is the $\sigma$-algebra of the measurable subsets of $M$. According to the previous requirements for memories and knowledge, $h$ must satisfy the following condition:
\begin{equation} \label{hi2}
\frac{\mu[H(t_2) \setminus H(t_1)]}{\mu[H(t_2)]} \leq \epsilon \ll 1 \; \hbox{for} \; t_1 \leq t_2 \leq t_1 + \Delta t,
\end{equation}
where $H(t)$ is the s-set $(t,h(t))$, and $\Delta t$ is a suitable non-infinitesimal amount of time. Condition (\ref{hi2}) guarantees that for any $t \in I$, the knowledge $h(t)$ can be remembered at least for a time $\Delta t$. A map $h$ satisfying (\ref{hi2}) will be referred to as a {\it branch}\footnote{This term is used here analogously to the quantum case, in which it is appropriate due to the tree structure of the universal wave function.}. One can say that branches represent the observable evolutions of the universe.

In order to simplify the mathematical formulation of the theory, hereafter we will assume $R^+$ as the time interval $I$, and $\Delta t = \infty$; the last equality express the assumption that knowledge must be remembered forever. With a reasoning analogous to that of the previous section, if $h$ is a branch, one can prove that for every $t$ the overwhelming majority of the paths belonging to $H(t)$ spent the overwhelming majority of the time interval $[0, t]$ inside the set $\bigcap_{ s \in [0, t]} H(s)$.

%newpage
\subsection{Path spaces and explanation: conclusion} \label{sec:conclusion}
In conclusion, a path space representing the universe explains the observed phenomena by defining: (1) the typical configurations at a fixed time relative to a subset of the configuration space, which explain the results of statistical experiments; (2) the branches, which explain the macroscopic evolution.

As shown by conditions (\ref{ex3}) and (\ref{hi2}), both these notions are defined by means of the set function
\begin{equation} \label{psf}
r_\mu(S_1|S_2):= \frac{\mu(S_2 \setminus S_1)}{\mu(S_2)}.
\end{equation}
This set function is only used in the typicality regime, i.e. when $r_\mu(S_1|S_2) \ll 1$, to define relative typicality. This means that $r_\mu(S_1|S_2) \ll 1$ implies that $S_1$ is typical relative to $S_2$, i.e. the overwhelming majority of the paths of $S_2$ also belong to $S_1$. For $t_1=t_2$, $r_\mu$ defines the typical configurations of a subset of the configuration space, while for $t_1 < t_2$ it constitutes the defining condition for branches. The set function $r_\mu$ will be referred to as the {\it probabilistic relative typicality function}.

Thus the only structure of a path space which is utilized in the explanation of the observed phenomena of the universe is the set function $r_\mu$ in the typicality regime, while the detailed structure of the probability measure and, of course, the structure of the paths, are empirically irrelevant.

%newpage
\section{Quantum Typicality Theory} \label{quantum}
In section \ref{path} we saw that the quantum formalism cannot define a probability measure on a set of paths. However, according to the results of the previous section, what we need in order to explain the observer phenomena is just the relative typicality function for s-sets. In this section we will show that the quantum formalism can provide such a function, without the need of an underlying probability measure. To our knowledge, a definition of typicality not based on a probability measure has never been explicitly proposed before in the literature, even if the possible independence of the two notions, probability and typicality, has been pointed out in \cite{goldstein}.

\subsection{Probabilistic typicality functions}
The first step is to study typicality more exactly in the probabilistic case. Let $(\Omega, {\cal F}, \mu)$ be a probability space. We have already seen the relative typicality function
\begin{equation}
r_\mu(A|B):=\frac{\mu(B \setminus A)}{\mu(B)}, \; A, B \in {\cal F},
\end{equation}
with the meaning $r_\mu(A|B) \ll 1$ implies that $A$ is typical relative to $B$, that is the overwhelming majority of the elements of $B$ also belong to $A$. It is useful to introduce two other typicality functions:
\begin{eqnarray}
 a_\mu(A)&:=&\mu(\overline{A}), \\
m_\mu(A,B)&:=&\frac{\mu(A \triangle B)}{\max\{\mu(A), \mu(B)\}}. \label{ditto}
\end{eqnarray}
The first one is the {\it absolute typicality function}, with the meaning $a_\mu(A) \ll 1$ implies that $A$ is typical relative to $\Omega$; the second one is the {\it mutual typicality function}, with the meaning $m_\mu(A,B) \ll 1$ implies that $A$ and $B$ are mutually typical, i.e. $A$ is typical relative to $B$ and vice-versa. The normalization factor of $m_\mu$ has been chosen from the following possibilities:
\begin{equation}
N_1=\max\{\mu(A), \mu(B)\}; \; N_2=[\mu(A) + \mu(B)]/2; \; N_3= \min\{\mu(A), \mu(B)\}.
\end{equation}
It is easy to show that, by defining $m^i_\mu:=\mu(A \triangle B)/N_i$, we have 
\begin{equation} \label{compi}
m^1_\mu \leq m^2_\mu \leq m^3_\mu \leq \frac{m^1_\mu}{1-m^1_\mu} \leq 2m^1_\mu \; \; \hbox{for} \; \; m^1_\mu \leq 0.5.
\end{equation}
The inequality $ m^3_\mu \leq m^1_\mu/(1-m^1_\mu)$ derives from the inequality $\mu(A \triangle B) \geq N_1 - N_3$. Thus the three set functions $m^i_\mu$ are equivalent in the typicality regime, i.e. $m^i_\mu(A,B) \ll1 \Leftrightarrow m^j_\mu(A,B) \ll1$ for any $i,j$. The normalization factor $N_1=\max\{\mu(A), \mu(B)\}$ has been chosen because, in this way, $a_\mu$ and $r_\mu$ can be expressed in terms of $m_\mu$. We have in fact:
\begin{eqnarray} 
a_\mu(A) & = & m_\mu(\Omega, A);\label{clit} \\
r_\mu(A|B) & = & m_\mu(A \cap B, B).
\end{eqnarray}
Note that the inequalities

\begin{equation}
m_\mu(A,B) \leq r_\mu(A|B) + r_\mu(B|A) \leq \frac{m_\mu(A,B)}{1 - m_\mu(A,B)} 
\end{equation}
guarantee the implication $r_\mu(B|A), r_\mu(B|A) \ll 1 \Leftrightarrow m_\mu(A, B) \ll 1$, which must hold for obvious reasons.

A last interesting set function is the following:
\begin{equation} \label{qqtt}
\tau_\mu(A, B):=\frac{2\mu(A \cap B)}{\mu(A) + \mu(B)}= 1 - \frac{\mu(A \triangle B)}{\mu(A) + \mu(B)}.
\end{equation}
We have that $0 \leq \tau_\mu(A, B) \leq 1$; $\tau_\mu(A, B)=0$ iff $\mu(A \cap B)=0$; $\tau_\mu(A, B)=1$ iff $\mu(A \triangle B)=0$; $\tau_\mu(A, B) \approx 1 \Leftrightarrow m_\mu(A, B) \ll 1$. Since these properties resemble those of a probability measure, the set function $\tau_\mu$ will be referred to as the probabilistic mutual typicality {\it measure}.

\subsection{The origin of quantum typicality} \label{sec:origin}

The quantum formalism allows us to define the single-time distribution of a stochastic process, namely $\mu_Q[(t, \Delta)]:=||E(\Delta)\Psi(t)||^2$, but, apparently, it does not provide any correlation between different time s-sets, because the finite dimensional distributions (\ref{notadd}) are not additive. However we argue that there is a kind of correlation between two different time s-sets which can be extracted from the quantum formalism, even if it is not as detailed as the finite dimensional distributions. This correlation is expressed in terms of mutual typicality, and it can be mathematically represented by means of a mutual typicality function analogous to (\ref{ditto}), but deriving from the quantum formalism.

The origin of such a correlation is based on a very natural assumption. Suppose that the wave function of a particle is the sum of two non-overlapping wave packets. The assumption is that, during the time in which the wave packets are non-overlapping, the particle stays inside the support of one of the two wave packets, without jumping to the other.

Let $\phi$ and $\phi_\perp:=\Psi(t_1)- \phi$ be the two wave packets at a time $t_1$, where $\Psi(t)$ is the wave function of the particle. At a time $t_2 > t_2$ the two wave packets will be $\phi(t_2):=U(t_2-t_1)\phi$ and $\phi_\perp(t_2):=U(t_2-t_1)\phi_\perp$, where $U(t)$ is the unitary time evolution operator. The requirement that the two wave packets are non-overlapping at the times $t_1$ and $t_2$ implies that there exist two subsets $\Delta_1$ and $\Delta_2$ of the configuration space of the particle such that
\begin{equation} \label{cc1}
\phi \approx E(\Delta_1)\Psi(t_1) \; \; \hbox{and} \; \; U(t_2-t_1)\phi \approx E(\Delta_2)\Psi(t_2),
\end{equation}
where $E(\cdot)$ is the projection-valued measure on the configuration space of the particle. The sets $\Delta_1$ and $\Delta_2$ can be considered as the supports of $\phi$ and $U(t_2-t_1)\phi$ respectively. The conditions (\ref{cc1}) can be combined to give the condition
\begin{equation} \label{cc2}
U(t_2-t_1)E(\Delta_1)\Psi(t_1) \approx E(\Delta_2)\Psi(t_2).
\end{equation}
This reasoning can also be reversed: given two subsets $\Delta_1$ and $\Delta_2$ satisfying condition (\ref{cc2}), the wave packet $\phi:= E(\Delta_1)\Psi(t_1)$ satisfies the conditions of (\ref{cc1}).

Therefore the condition $||E(\Delta_2)\Psi(t_2) - U(t_2-t_1)E(\Delta_1)\Psi(t_1)|| \approx 0$, properly normalized, implies that a trajectory belonging to $(t_1, \Delta_1)$ also belongs (almost certainly) to $(t_2, \Delta_2)$, and vice-versa, i.e. that the two s-sets $(t_1, \Delta_1)$ and $(t_2, \Delta_2)$ are mutually typical. This result will be formalized in the next subsections.

\subsection{Quantum typicality functions} \label{sec:qtm}
Consider the space $(M^T, {\cal S})$, and assume as usual that a normalized universal wave function $\Psi(t)=U(t)\Psi_0$ is given. In order to simplify the notation, given $S=(t, \Delta) \in {\cal S}$, let $S\Psi_0$ denote the state $U^\dagger(t)E(\Delta)U(t)\Psi_0$.

 Let us define the {\it quantum mutual typicality function} as
\begin{equation} \label{uyt}
m_\Psi(S_1, S_2):=\frac{||S_1\Psi_0 - S_2 \Psi_0||^2} {\max\{ ||S_1 \Psi_0||^2, ||S_2 \Psi_0||^2\}}.
\end{equation}
An explicit expression for (\ref{uyt}) is 
\begin{equation} \label{seri}
m_\Psi(S_1, S_2)=\frac{||E(\Delta_2)\Psi(t_2) - U(t_2 - t_1)E(\Delta_1) \Psi(t_1)||^2}{\max\{ ||E(\Delta_1)\Psi(t_1)||^2, ||E(\Delta_2)\Psi(t_2)||^2\}},
\end{equation}
another possibility being the same expression with 1 and 2 interchanged. Thus we see that the definition (\ref{uyt}) corresponds to the typicality function introduced in the previous subsection. Note that $m_\Psi$ is defined on ${\cal S} \times {\cal S}$ and not on $\sigma({\cal S}) \times \sigma({\cal S})$, as in the probabilistic case. Here too the chosen normalization factor is $\max\{ ||S_1 \Psi_0||^2, ||S_2 \Psi_0||^2\}$. Other possible normalization factors are defined analogously to the probabilistic case, and the inequalities (\ref{compi}) become
\begin{equation} \label{rup}
m^1_\Psi \leq m^2_\Psi \leq m^3_\Psi \leq \frac{m^1_\Psi}{\left (1-\sqrt{m^1_\Psi}\right )^2} \leq 2 m^1_\Psi \; \; \hbox{for} \; \; m^1_\Psi \leq 0.08,
\end{equation}
in such a way that, also in the quantum case, the different normalization factors are equivalent in the typicality regime. 

For  two equal time s-sets $S_1=(t, \Delta_1)$ and $S_2=(t, \Delta_2)$, the function $m_\Psi$ becomes 
\begin{equation} \label{bt1}
m_\Psi(S_1, S_2)=\frac{||E(\Delta_1 \triangle \Delta_2)\Psi(t)||^2} {\max\{||E(\Delta_1)\Psi(t)||^2, ||E(\Delta_2)\Psi(t)||^2\}},
\end{equation}
which is the probabilistic mutual typicality function deriving from the probability measure $||E(\cdot)\Psi(t)||^2$. 

In order to interpret (\ref{uyt}) as a function defining typicality, one must postulate the following
\begin{trivlist}
\item[\hspace\labelsep{\bf Quantum Typicality Rule:}]
\begin{equation}
m_\Psi(S_1, S_2) \ll 1 \Rightarrow S_1 \, \hbox{and} \, S_2 \, \hbox{are mutually typical}.
\end{equation}
\end{trivlist}
There is a strong analogy between this rule and the Born rule, as we will see better at the end of this section. The main consequence of the quantum typicality rule is that the typical trajectories of the universe follow the branches of the universal wave function, as we will see in section \ref{sec:subtree}. Another way to look at this rule is related to information: if $m_\Psi(S_1, S_2) \ll 1$, the information that the trajectory of the universe was inside $\Delta_1$ at the time $t_1$ is not lost at the time $t_2$. 

By analogy with equation (\ref{clit}), we define the {\it quantum absolute typicality function} as:
\begin{equation} \label{bt2}
a_\Psi(S):=m_\Psi(S, M^T)=||S\Psi_0- \Psi_0||^2=||E(\overline{\Delta})\Psi(t)||^2.
\end{equation}
Note that (\ref{bt2}) is equal to the probabilistic absolute typicality function deriving from the probability measure $||E(\cdot)\Psi(t)||^2$.

As to the {\it quantum relative typicality function} $r_\Psi(S_1|S_2):=m_\Psi(S_1 \cap S_2, S_2)$, since $S_1 \cap S_2 \in {\cal S}$ only if $S_1$ and $S_2$ are equal time s-sets, it is defined only in that case. Thus, given $S_1=(t, \Delta_1)$ and $S_2=(t, \Delta_2)$ we have:
\begin{equation} \label{bt3}
r_\Psi(S_1|S_2):=m_\Psi(S_1 \cap S_2, S_2)=\frac{||(t, \Delta_1 \cap \Delta_2)\Psi_0 - S_2\Psi_0||^2} {||S_2\Psi_0||^2} = \frac{||E(\Delta_2 \setminus \Delta_1)\Psi(t)||^2} {||E(\Delta_2)\Psi(t)||^2}.
\end{equation}
Again, for equal time s-sets, $r_\Psi$ is equal to the probabilistic relative typicality function deriving from $||E(\cdot)\Psi(t)||^2$.

Even if the quantum formalism does not allow directly defining $r_\Psi(S_1|S_2)$ when $t_1 \neq t_2$, it is possible to provide an indirect definition for such a function. Consider the lower bound
\begin{equation} \label{potr}
\inf_{\Delta \in {\cal B}} ||S_1\Psi_0 - (t_2, \Delta)\Psi_0||=\inf_{\Delta \in {\cal B}} ||E(\Delta)\Psi(t_2)- U(t_2-t_1)E(\Delta_1)\Psi(t_1)||.
\end{equation}
It is a minimum, and the natural set $\tilde{\Delta}$ corresponding to the minimum is
\begin{equation}
\tilde{\Delta}=\{x \in M: |\langle x |\Psi(t_2) \rangle|^2 < 2Re \langle \Psi(t_2)| x \rangle \langle x |U(t_2-t_1) E(\Delta_1) |\Psi(t_1) \rangle \}.
\end{equation}
This can be seen by inserting the identity $I=\int |x\rangle dx \langle x|$ into the scalar products. Thus we can define $r_\Psi(S_1|S_2)$ as:
\begin{equation} \label{rtyu}
r_\Psi(S_1|S_2):=\max\{m_\Psi[S_2 \cap (t_2, \tilde{\Delta}), S_2], m_\Psi[S_1, (t_2, \tilde{\Delta})]\}.
\end{equation}
If $r_\Psi(S_1|S_2) \ll 1$, both functions in the right hand member of (\ref{rtyu}) are $\ll 1$. Thus, according to the first term, the overwhelming majority of the trajectories of $S_2$ belong to $(t_2, \tilde{\Delta})$, and according to the second term the overwhelming majority of the trajectories of $(t_2, \tilde{\Delta})$ belong to $S_1$. As a consequence, $S_1$ is typical relative to $S_2$. Note however that this function may fail to work when $||S_2\Psi_0||$ is too small, that is when $||S_2\Psi_0|| \approx ||S_1\Psi_0- (t_2, \tilde{\Delta})\Psi_0||$.

The quantum mutual typicality {\it measure} $\tau_\Psi$ is defined analogously to (\ref{qqtt}):
\begin{equation}
\tau_\Psi(S_1, S_2):=\frac{2 |Re \langle \Psi_0 | S_1 S_2 |\Psi_0 \rangle|}{||S_1\Psi_0||^2 + ||S_2\Psi_0||^2}=\left |1 - \frac{||S_1\Psi_0 - S_2\Psi_0||^2}{||S_1\Psi_0||^2 + ||S_2\Psi_0||^2} \right |.
\end{equation}
We have: $0 \leq \tau_\Psi(S_1, S_2) \leq 1$; $\tau_\Psi(S_1, S_2) = 0 $ iff $Re \langle \Psi_0 | S_1 S_2 |\Psi_0 \rangle =0$; $\tau_\Psi(S_1, S_2) = 1 $ iff $S_1\Psi_0=S_2\Psi_0$; $\tau_\Psi(S_1, S_2) \approx 1 \Leftrightarrow m_\Psi(S_1, S_2) \ll 1$. 

\vspace{3mm}
With respect to typicality, the quantum typicality rule plays the same conceptual role that the Born rule plays with respect to probability. Actually, the quantum typicality rule is the extension to unequal time s-sets of the Born rule in the typicality regime. Indeed, for equal time s-sets, all the quantum typicality functions, namely (\ref{bt1}), (\ref{bt2}), (\ref{bt3}), have the same form and the same meaning of the corresponding probabilistic typicality functions obtained from the Born rule, i.e. assuming that $||E(\cdot )\Psi(t)||^2$ is a probability measure. On the contrary the Born rule has nothing to say about the mutual typicality of non equal time s-sets. Actually, one could try to define mutual typicality by means of an expression of the type:
\begin{equation}
m_\Psi(S_1, S_2):=
\frac{||E(\overline{\Delta}_2)U(t_2-t_1)E(\Delta_1)\Psi(t_1)||^2}{||E(\Delta_1)\Psi(t_1)||^2}+ \frac{||E(\overline{\Delta}_1)U(t_1-t_2)E(\Delta_2)\Psi(t_2)||^2}{||E(\Delta_2)\Psi(t_2)||^2}.
\end{equation}
According to the Born rule, if $t_2 \geq t_1$, the first term is the probability that a trajectory belonging to $\Delta_1$ at the time $t_1$ belongs to $\overline{\Delta}_2$ at the time $t_2$. By assuming a sort of reverse Born rule, the same meaning (with 1 and 2 interchanged) can be given to the second term. However, this definition is surely less natural and more complex than definition (\ref{uyt}).

One last remark about the definition of the quantum typicality function $m_\Psi$: Due to the vagueness of the notion of typicality, the set function $M_\Psi(S_1, S_2):=\sqrt{m_\Psi(S_1, S_2)}$ could also be a possible definition for the mutual typicality function. The definition $m_\Psi$ has the advantage that, for equal time s-sets, it reduces to the typicality function deriving from the Born rule. On the other hand, the definition $M_\Psi$ has the advantage that the proof of some consistency conditions is more simple, due to the fact that $||S_1\Psi_0 - S_2\Psi_0||$ is a distance. Further studies may suggest adopting  $M_\Psi$ instead of $m_\Psi$ as the definition of the quantum mutual typicality function.

\subsection{Typicality function and non-overlapping wave packets} \label{sec:over}

In this subsection we study the connection between the quantum mutual typicality function and the non-overlapping property of the wave packets. Due to the spreading of the wave packets, such a property must be considered in an approximate way; appropriate mathematical tools will be developed to this purpose.

\vspace{3mm}
Given a state $\phi \in L^2(M)$, we say that $\Delta \in {\cal B}$ is {\it a support} for $\phi$ if
\begin{equation}
\frac{||\phi - E(\Delta)\phi||^2}{||\phi||^2} \ll 1.
\end{equation}
The overlapping degree of two states $\phi_1, \phi_2 \in L^2(M)$ can be expressed by the following {\it overlapping measure}:
\begin{equation} \label{nop}
w(\phi_1, \phi_2):= \inf_{\Delta \in {\cal B}} \frac{||E(\overline{\Delta})\phi_1||^2 + ||E(\Delta)\phi_2||^2}{\min\{||\phi_1||^2, ||\phi_2||^2\}}=\frac{\int{ \min\{|\phi_1(x)|^2, |\phi_2(x)|^2\} dx}}{\min\{||\phi_1||^2, ||\phi_2||^2\}}.
\end{equation}
Note that 
$$
||E(\overline{\Delta})\phi_1||^2 + ||E(\Delta)\phi_2||^2=||E(\overline{\Delta})\phi_1 \pm E(\Delta)\phi_2||^2=||\phi_1 - E(\Delta)(\phi_1 + \phi_2)||^2.
$$
We have $0 \leq w(\phi_1, \phi_2) \leq 1$; $w(\phi_1, \phi_2)=0$ iff $\phi_1(x)\phi_2(x)=0$ almost everywhere, and $w(\phi_1, \phi_2)=1$ iff $|\phi_1(x)| \leq |\phi_2(x)|$ or $|\phi_1(x)| \geq |\phi_2(x)|$ almost everywhere. The expression of the lower bound (\ref{nop}) assumes its minimum value for the set
\begin{equation}
\tilde{\Delta}:=\{x \in M : |\phi_1(x)| > |\phi_2(x)| \}.
\end{equation}
If $w(\phi_1, \phi_2) \ll 1$ than  $\phi_1$ and $\phi_2$ admit disjoined supports, and therefore we say that they are {\it non-overlapping}. 

\vspace{3mm}
Let us study now the wave packets of the universal wave function. Given an s-sets $S_1=(t_1, \Delta_1)$, with $||S_1\Psi_0||^2 \leq 1/2$, let us consider the states $U(t_2)S_1\Psi_0=U(t_2-t_1)E(\Delta_1)\Psi(t_1)$ and $U(t_2)\bar{S}_1\Psi_0=\Psi(t_2)- U(t_2)S_1\Psi_0$. The overlapping measure of the two states is:
\begin{equation} \label{terro}
w [U(t_2)S_1\Psi_0, U(t_2)\bar{S}_1\Psi_0]=\frac{\inf_{\Delta_2}||S_1\Psi_0 - S_2\Psi_0||^2}{||S_1\Psi_0||^2}=\frac{ ||S_1\Psi_0 - (t_2, \tilde{\Delta}_2) \Psi_0||^2}{||S_1\Psi_0||^2},
\end{equation}
where $S_2=(t_2, \Delta_2)$, and 
\begin{equation}
\tilde{\Delta}_2:=\{ x \in M : | \langle x |\Psi(t_2) \rangle |^2 < 2Re \langle \Psi(t_2)|x \rangle \langle x |U (t_2)S_1|\Psi_0 \rangle\}.
\end{equation}
From (\ref{terro}) we obtain the following inequalities:
\begin{equation} \label{ovart}
\inf_{\Delta_2} m_\Psi(S_1, S_2) \leq w [U(t_2)S_1\Psi_0, U(t_2)\bar{S}_1\Psi_0] \leq \inf_{\Delta_2} m_\Psi^3(S_1, S_2),
\end{equation}
where
$$
m^3_\Psi(S_1, S_2) := \frac{||S_1\Psi_0 - S_2\Psi_0||^2}{\min\{||S_1\Psi_0||^2, ||S_2\Psi_0||^2\}}.
$$
Moreover, we have
\begin{equation} \label{ovart2}
\frac{||U(t_2)S_1\Psi_0- E(\Delta_2)U(t_2)S_1\Psi_0||^2}{||S_1\Psi_0||^2} \leq m^3_\Psi(S_1, S_2).
\end{equation}
Since $m^3_\Psi(S_1, S_2) \ll 1 \Leftrightarrow m_\Psi(S_1, S_2) \ll 1$ (inequalities (\ref{rup})), from the inequalities (\ref{ovart}) and (\ref{ovart2}) we obtain the implications
\begin{equation} \label{ii1}
w(U(t_2)S_1\Psi_0, U(t_2)\bar{S}_1\Psi_0) \ll 1 \Leftrightarrow \inf_{\Delta_2} m_\Psi(S_1, S_2) \ll 1,
\end{equation}
\begin{equation} \label{ii2}
m_\Psi(S_1, S_2) \ll 1 \Rightarrow \Delta_2 \; \; \hbox{is a support of} \; U(t_2)S_1\Psi_0,
\end{equation}
which express the relationship between the typicality function and the overlapping of the wave packets. In words, the first implication states that if there exists $\Delta_2$ such that $m_\Psi(S_1, S_2) \ll 1$, than the two wave packets $U(t_2)S_1\Psi_0$ and $U(t_2)\bar{S}_1\Psi_0$ are non-overlapping, and vice-versa.

\subsection{Asymptotic extension}
It is possible to extend the quantum typicality rule and the related formalism to the limit $t=\infty$.

Given a trajectory $\lambda \in M^T$, the limit 
\begin{equation}
v^+(\lambda):= \lim_{t \rightarrow +\infty} \frac{\lambda(t)}{t},
\end{equation}
if this exists, is referred to as the {\it asymptotic velocity} of $\lambda$. For instance, under very general assumptions for the Hamiltonian, one can prove that the trajectories of a classical system admit an asymptotic velocity \cite{scattering}, p. 245. Let $\tilde{M}^T$ denote the subset of $M^T$ composed of the trajectories admitting  the asymptotic velocity. Given $\Delta_v \subseteq R^{3N}$, let us define the {\it asymptotic} s-set $(\infty, \Delta_v)$ as
\begin{equation}
(\infty, \Delta_v) :=\{\lambda \in \tilde{M}^T : v^+(\lambda) \in \Delta_v\}.
\end{equation}
Let ${\cal A}$ denote the class of asymptotic s-sets, and let ${\cal C}:={\cal S} \cup {\cal A}$. We replace the space $(M^T, {\cal S})$ utilized in the previous section with the space $(\tilde{M}^T, {\cal C})$. With this replacement we assume that the admissible trajectories of the universe must have a well defined asymptotic velocity.

\vspace{3mm}
As to the quantum formalism, under very general assumption for the quantum Hamiltonian $H$, the limits 
\begin{equation}
\label{asy}
V_i^+:=s-\lim_{t \rightarrow +\infty}V_i^t:=s-\lim_{t \rightarrow +\infty}\frac{U^\dagger(t)Q_iU(t)}{t}, \; \; \hbox{for} \; \; i=1, \ldots, 3N
\end{equation}
do exist, where $Q_i$ are the position operators for the particles. The operators $\{V_i^+\}$ are referred to as the {\it asymptotic velocity operators}, and they commute with each others and with the Hamiltonian \cite{scattering}, p. 299.

Let us study the limit (\ref{asy}).  For a single particle whose Hamiltonian admits the wave operator $\Omega_+$, we have \cite{scattering}, p. 166:
\begin{equation}
{\bf V}^+=\Omega_+ \frac{\bf P}{m} \Omega_+^\dagger,
\end{equation}
Consider however that the asymptotic velocity operators exist even if the wave operator does not exist. For a free particle we have ${\bf V}^+={\bf P}/m$.

Let $E_x$, $F_v^t$ and $F_v^+$ denote the spectral families of $Q$, $V^t$ and $V^+$ respectively, and $E(\cdot)$, $F^t(\cdot)$ and $F^+(\cdot)$ their spectral measures (for simplicity, the coordinate-particle indices $i$ are omitted here). From the equalities
\begin{equation}
\int v dF_v^t = \int \frac{x}{t}U^\dagger(t) dE_x U(t)=\int vU^\dagger(t) dE_{vt} U(t),
\end{equation}
we obtain $F_v^t=U^\dagger(t) E_{vt} U(t)$, and $F^t(\Delta_v)=U^\dagger(t) E(t \Delta_v) U(t)$, where $t \Delta_v := \{vt \in M : v \in \Delta_v \}$. From the theory of convergence of the self-adjoint operators \cite{reed}, if $\partial \Delta_v$ does not belong to the pure point spectrum of $V^+$, one obtains 
\begin{equation}
s-\lim_{t \rightarrow +\infty} U^\dagger(t) E(t \Delta_v) U(t)=s-\lim_{t \rightarrow +\infty} F^t(\Delta_v)= F^+(\Delta_v).
\end{equation}

\vspace{3mm}
We can extend the quantum formalism of the previous subsections to $(\tilde{M}^T, {\cal C})$. Given $C \in {\cal C}$, let us define
\begin{equation}
C\Psi_0:= \left \langle
\begin{array}{ll}
U^\dagger(t)E(\Delta)U(t)\Psi_0 & \hbox{for} \; C =(t, \Delta) \in {\cal S}, \\
F^+(\Delta_v)\Psi_0 & \hbox{for} \; C = (\infty, \Delta_v) \in {\cal A}.
\end{array}\right.\nonumber 
\end{equation}
In this way, all the quantum typicality functions and the quantum typicality rule can be extended to $(\tilde{M}^T, {\cal C})$. For instance, given $S \in {\cal S}$ and $A=(\infty, \Delta_v) \in {\cal A}$, if $m_\Psi(S, A) \ll 1$ then the overwhelming majority of the trajectories belonging to $S$ have an asymptotic velocity belonging to $\Delta_v$, and vice versa.

\vspace{3mm}
The results obtained in subsection \ref{sec:over} relative to the wave packet $U(t)S_1\Psi_0$ also hold for a wave packet of the type $U(t)C\Psi_0$, where $C \in {\cal C}$, and always with $||C\Psi_0||^2 \leq 1/2$. Moreover, it is possible to calculate the limit $w[U(t)C\Psi_0, U(t)\bar{C}\Psi_0]$ for $t \rightarrow +\infty$. Let us consider indeed the lower bound $\inf_{\Delta_v} ||C\Psi_0 - F^+(\Delta_v)\Psi_0||$. The minimum value is reached for the set
\begin{equation} 
\tilde{\Delta}_v:=\{ v \in R^{3N}:  \sum_{\alpha_v} |\langle v, \alpha_v |\Psi_0 \rangle|^2 < \sum_{\alpha_v} 2Re \langle \Psi_0|\alpha_v, v \rangle \langle v, \alpha_v |C|\Psi_0 \rangle\},
\end{equation}

where $\{|\alpha_v, v \rangle\}$ is a complete set of generalized eigenvectors of the asymptotic velocities ($\alpha_v$ being the quantum numbers resolving the possible degeneracy of the eigenvalue $v$). We have
\begin{equation} \label{toprov}
\lim_{t \rightarrow +\infty} \inf_\Delta ||C\Psi_0 -  (t, \Delta)\Psi_0||=||C\Psi_0 -  F^+(\tilde{\Delta}_v)\Psi_0||.
\end{equation}
Indeed $\inf_\Delta||C\Psi_0 - (t, \Delta)\Psi_0||=\inf_{\Delta_v}||C\Psi_0 - F^t(\Delta_v)\Psi_0||$, and
\begin{eqnarray*} 
& & \left | \inf_{\Delta_v} ||C\Psi_0 - F^t(\Delta_v)\Psi_0|| - ||C\Psi_0 - F^+(\tilde{\Delta}_v)\Psi_0||\right | \leq \\
& & \inf_{\Delta_v} || F^t(\Delta_v)\Psi_0 - F^+(\tilde{\Delta}_v)\Psi_0|| \leq  || F^t(\tilde{\Delta}_v)\Psi_0 - F^+(\tilde{\Delta}_v)\Psi_0|| \rightarrow 0 \; \; \hbox{for} \; \; t \rightarrow \infty.
\end{eqnarray*}
Thus
\begin{equation}
\lim_{t \rightarrow +\infty} w[U(t)C\Psi_0, U(t)\bar{C}\Psi_0] = \frac{||C\Psi_0 - F^+(\tilde{\Delta}_v)\Psi_0||^2}{||C\Psi_0||^2},
\end{equation}
and the equation ({\ref{ovart}) is valid also at the time $t_2=+\infty$. In this case the set $S_2$ becomes an asymptotic s-set.

\subsection{Subtrees and branches} \label{sec:subtree}

In this section the mathematical definitions of subtrees and branches as non-overlapping parts of the universal wave function are given.

Branches are present, in a more or less explicit manner, in many formulations of quantum mechanics, namely Bohmian mechanics, the Many Worlds Interpretation \cite{everett, mwi}, the Consistent Histories formulation of quantum mechanics \cite{consistent} and the theory of decoherence \cite{decoherence}. The definition of branches as non-overlapping parts of the universal wave function is present mainly in the works connected with Bohmian mechanics, for instance \cite{bohm3, struyve, peruzzi, deotto}. According to these authors, during its evolution the universal wave function splits into permanently non-overlapping wave packets, for instance in the presence of a measurement. This process is also called the {\it effective collapse} of the wave function. Here a schematic description of the process. 

Let us suppose that during the time interval $(t_1, t_2)$ a measurement with two possible outcomes is performed on a quantum system. At the time $t_1$  the wave function of the universe is of the form $\Psi(t_1)=(\varphi_+ +  \varphi_-) \otimes \Phi_0 \otimes \Psi_E(t_1)$, where $\varphi_\pm$ are eigestates of the quantum system corresponding to the measured observable, $\Phi_0$ is the state of the measuring device before the measurement, and $\Psi_E(t_1)$ is the state of the environment, i.e. of the rest of the universe. At the time $t_2$, when the measurement has been just performed, the universal wave function is of the form $\Psi(t_2)=(\varphi_+ \otimes \Phi_+ +  \varphi_- \otimes \Phi_-) \otimes \Psi_E(t_2)$, where $\Phi_+$ and $\Phi_-$ are the states of the measuring device which has recorded the results $+$ and $-$ respectively. Since $\Phi_+$ and $\Phi_-$ represent the instrument with a pointer in two macroscopically distinct positions, they are non-overlapping. The measuring device unavoidably interacts with the environment; thus, at a subsequent time $t_3$, we have $\Psi(t_3)=\varphi_+ \otimes \Phi_+ \otimes \Psi_E^+(t_3) + \varphi_- \otimes \Phi_- \otimes \Psi_E^-(t_3)$, where $\Psi_E^+(t_3)$ and $\Psi_E^-(t_3)$ are the states of the environment which have interacted with $\Phi_+$ and $\Phi_-$ respectively. It is easy to accept that $\Psi_E^+(t_2)$ and $\Psi_E^-(t_2)$ are permanently non-overlapping: remember that it is sufficient that a single particle has two different positions in $\Psi_E^+(t_2)$ and $\Psi_E^-(t_2)$ in order to guarantee that the two states are non-overlapping. Of course, the splitting of the universal wave function in permanently non-overlapping wave packets may occur in many other different situations, non only during a measurement.

\vspace{3mm}
This is the usual semi-qualitative description of the branching process of the universal wave function.  We propose now an explicit definition for the branches, which is based on the mathematical formalism developed in the previous sections. 

The first step is to define the subtree-supports. We say that an s-set $S_1=(t_1, \Delta_1)$ is a {\it (forward) subtree-support} if $||S_1 \Psi_0||^2 \leq 1/2$, and moreover the states $U(t) S_1\Psi_0=U(t-t_1)E(\Delta_1)\Psi(t_1)$ and $ U(t) \bar{S}_1\Psi_0=\Psi(t) - U(t) S_1\Psi_0$ are non-overlapping for $t \geq t_1$. In mathematical terms:
\begin{equation} \label{lobb2}
w[U(t)S_1\Psi_0, U(t)\bar{S}_1\Psi_0] \leq \epsilon \ll 1 \; \; \hbox{for} \; \; t \geq t_1,
\end{equation}
where $w$ is the overlapping measure defined by (\ref{nop}). 
Given the above definition of subtree-support, it is natural to define a {\it (forward) subtree} as a mapping $k:[t_0, +\infty) \rightarrow {\cal B}$ satisfying the condition
\begin{equation} \label{subt1}
m_\Psi[K(t_1), K(t_2)] \leq \epsilon \ll 1 \; \; \hbox{for} \; \; t_0 \leq t_1, t_2 < \infty,
\end{equation}
where $K(t):=(t, k(t)) \in {\cal S}$. According to the implications (\ref{ii1}) and (\ref{ii2}), this condition guarantees that $K(t)$ is a subtree-support for every $t \geq t_0$, and that, for $t_1, t_2 \geq t_0$, the set $k(t_2)$ is a support of the state $U(t_2-t_1)E[k(t_1)]\Psi(t_1)$. Moreover, according to the quantum typicality rule, for any $t_1, t_2 \geq t_0$ the overwhelming majority of the trajectories of $K(t_1)$ also belong to $K(t_2)$, and vice-versa. If the definition of $K(t)$ derived from a mutual typicality measure of probabilistic nature, with a reasoning analogous to that of section \ref{explanation} one could deduce that for any time $t_1 \geq t_0$ and for the overwhelming majority of the times $t_2 \geq t_0$, the overwhelming majority of the trajectories belonging to $K(t_1)$ also belong to $K(t_2)$. Arguably such a conclusion can be extended to the case in which the typicality measure is of a quantum nature, even if this extension would have to be supported by further studies on the interpretation of typicality. The conclusion is that the trajectories of the particles follow approximately the subtrees of the universal wave function.

It is useful to introduce the notion of asymptotic subtree-support: we say that an s-set $S_1$  is an {\it asymptotic} subtree-support if the states $U(t) S_1\Psi_0$ and $ U(t) \bar{S}_1\Psi_0$ are non-overlapping at the time $t=+\infty$, that is:
\begin{equation}
\lim_{t \rightarrow +\infty} w[U(t)S_1\Psi_0, U(t)\bar{S}_1\Psi_0] \ll 1.
\end{equation}
Therefore the two states may sometimes overlap in the time interval $(t_1, \infty)$; however the information that at the time $t_1$ the trajectory was in $\Delta_1$ is not lost, and it can be recovered at least at the time $+\infty$. Of course a subtree-support is also an asymptotic subtree-support, but the contrary is not true. Consider for instance a particle in one dimension, whose initial wave function $\Psi_0$ is the sum of two non-overlapping Gaussian wave packets $\phi_\pm$, with mean positions $\pm|x_0|$ and mean momenta $\mp|p_0|$. The two wave packets move in opposite directions, overlap in the neighbourhood of the origin and then move away and become permanently non-overlapping. The s-sets $(0, R^\pm)$ are asymptotic subtree-supports but not subtree-supports. This example allows us to show an important difference between the trajectories defined by the quantum typicality rule and those defined by Bohmian mechanics. Since Bohmian trajectories cannot cross each other, in this example Bohmian trajectories belonging for instance to $(0,R^-)$ ``bounce'' and belong to $(t,R^-)$ for every $t$. On the other hand, according to the quantum typicality rule, the overwhelming majority of the trajectories belonging to $(0,R^-)$ will belong to $(t,R^+)$ after a suitable time $t_0$.

Another meaningful definition is that of an irreducible subtree-support. We say that an s-set $S=(t, \Delta)$ is an {\it irreducible subtree-support} if it is an asymptotic subtree-support, and moreover, for any other asymptotic subtree-support $S'=(t, \Delta') \subseteq S$, we have
\begin{equation}
m_\Psi(S, S') \ll 1 \; \; \hbox{and} \; \; \frac{\mu_L(\Delta \triangle \Delta')}{\max\{\mu_L(\Delta), \mu_L(\Delta')\}} \ll 1,
\end{equation}
where $\mu_L$ is the Lebesgue measure on $M$. In words, $S$ does not ``properly" contain any asymptotic subtree-support, and its spatial extension is the minimum extension compatible with being an asymptotic subtree-support. The information that the trajectory of the universe is in some proper subset of an irreducible subtree-support is destined to be lost, because after a suitable time there is no longer any spatial measurement which can recover such information. This is the case, for instance, with the two-slit experiment, in which the information of the slit crossed by the particle is definitively lost when the two wave packets emerging from the slits overlap and hit the screen. If we assume, as in section \ref{sec:memory}, that what can be known is only what can be remembered forever, then for no observer can the knowledge of the position of the trajectory of the universe exceed the knowledge represented by an irreducible subtree-support.

\vspace{3mm}

By using the relative typicality function (\ref{rtyu}) we can define branches: a mapping $h:[t_0, +\infty) \rightarrow {\cal B}$ is a {\it branch} if $||H(t)\Psi_0||^2 \leq 1/2$ for $t \in [t_0, +\infty)$, where $H(t):=(t, h(t))$, and moreover
\begin{equation} \label{bra1}
r_\Psi[H(t_1)|H(t_2)] \leq \epsilon \ll 1 \; \; \hbox{for} \; \; t_0 \leq t_1 \leq t_2.
\end{equation}
Due to the structure of $r_\Psi$, every s-set $H(t)$ is also a subtree-support. According to the meaning of $r_\Psi$, the branches have the required property relative to typicality, i.e. $t_2 \geq t_1$ implies that $H(t_1)$ is typical relative to $H(t_2)$.

\vspace{3mm}
Two last remarks. The definitions of subtrees and branches are {\it vague}, that is no definite value for $\epsilon$ in (\ref{subt1}) and (\ref{bra1}) is given. Moreover, probably it is possible to give other equivalent definitions for such entities. However this is not a problem, because subtrees and branches are not structural elements of this formulation, but rather descriptions of the influence of the universal wave function on the trajectories. Note that this is not the case in the Many Worlds Interpretation, where the branches, i.e. the worlds, constitute the primitive ontology of that interpretation, and the vagueness of their definition is surely a problem.

Since the overwhelming majority of the trajectories follow the branches of the universal wave function, this formulation of quantum mechanics explains the quasi-classical macroscopic evolution of the universe only if the universal wave function actually has a branch structure, and if the branches have a quasi-classical structure. Here we do not face the problem of proving this, and we limit ourselves to the argument that the Ehrenfest theorem and Mott's analysis of the cloud chamber \cite{mott} should be important tools to obtain more rigorous results in this sense.

\subsection{On the consistency of the quantum typicality rule}
In order to guarantee that the quantum typicality rule is consistent, the quantum mutual typicality function must reflect the structural properties of mutual typicality. For instance, we cannot have $m_\Psi(S_1, S_2), m_\Psi(S'_1, S_2) \ll 1$ and $S_1 \cap S'_1= \emptyset$ at the same time. In this subsection we present some inequalities satisfied by the quantum typicality function, which guarantee that some natural structural properties of mutual typicality are satisfied.

For the mutual quantum typicality function $m$ (in this subsection the subscript $\Psi$ will be omitted) we have the following inequalities:
\begin{eqnarray}
& & m(S_1, S_3) \leq m^3(S_1, S_2) + m^3(S_2, S_3) + 2\sqrt{m^3(S_1, S_2)m^3(S_2, S_3)}; \label{ine1}  \\
& & m(S_1 \cap S'_1, S_2), m(S_1 \cup S'_1, S_2) \leq m^3(S_1, S_2) + m^3(S'_1, S_2); \label{ine2} \\
& & 1 - w(S, S') \leq \frac{1}{2} m^3(S, S'); \label{ine3} \\
& & w(S_2, S_2') \leq a \sqrt{m(S_1, S_2)}+b \sqrt{m(S'_1, S'_2)} + c w(S_1, S'_1), \label{ine4}
\end{eqnarray}
where: $S$ and $S'$ in (\ref{ine3}), and $S_i$ and $S'_i$, $i=1, 2, \ldots$ in (\ref{ine2}) and (\ref{ine4}) are equal time s-sets;
\begin{eqnarray*}
& & a = \frac{\max\{||S_1\Psi_0||,||S_2\Psi_0||\} ||S'_2\Psi_0||}{\min\{||S_2\Psi_0||^2,||S'_2\Psi_0||^2\}}, \; \; 
b = \frac{\max\{||S'_1\Psi_0||,||S'_2\Psi_0||\} ||S_1\Psi_0||}{\min\{||S_2\Psi_0||^2,||S'_2\Psi_0||^2\}}, \\
& & c = \frac{\min\{||S_1\Psi_0||^2,||S'_1\Psi_0||^2\}} {\min\{||S_2\Psi_0||^2,||S'_2\Psi_0||^2\}};
\end{eqnarray*}
$w(S, S')$ is a shorthand notation for $w[E(\Delta)\Psi(t), E(\Delta')\Psi(t)]$. Note that
$$
w(S, S')=\frac{||(S \cap S')\Psi_0||^2} {\min\{||S\Psi_0||^2,||S'\Psi_0||^2\}}.
$$
{\bf Proof.} The inequality (\ref{ine1}) derives from the triangle inequality $||S_1\Psi_0 - S_3\Psi_0|| \leq ||S_1\Psi_0 - S_2\Psi_0|| + ||S_2\Psi_0 - S_3\Psi_0||$. Inequalities (\ref{ine2}) derive from the equality
$$
||(S_1 \cup S_1')\Psi_0 - S_2\Psi_0||^2 + ||(S_1\cap S'_1)\Psi_0 - S_2\Psi_0||^2 = 
||S_1\Psi_0 - S_2\Psi_0||^2 + ||S_1'\Psi_0 - S_2\Psi_0||^2. 
$$

Inequality (\ref{ine3}) is straightforward. Inequality (\ref{ine4}) is obtained by applying the Schwarz inequality to the right hand member of the equation
$$
\langle \Psi_0| S_2 S_2'| \Psi_0 \rangle= \langle \Psi_0| (S_2 -S_1)S_2' + S_1(S_2' - S_1') + S_1S_1'| \Psi_0 \rangle,
$$
and then slightly manipulating. {\bf q.e.d.} 

\vspace{3mm}
From inequalities (\ref{ine1}) to (\ref{ine4}) we obtain the following implications:
\begin{eqnarray}
& & m(S_1, S_2), m(S_2, S_3) \ll 1 \Rightarrow m(S_1, S_3) \ll 1; \label{impli1} \\
& & m(S_1, S_2), m(S'_1, S_2) \ll 1 \Rightarrow m(S_1 \cap S'_1, S_2), m(S_1 \cup S'_1, S_2) \ll 1; \label{impli2} \\
& & m(S_1, S_2), m(S'_1, S_2) \ll 1 \Rightarrow  1 - w(S_1, S'_1) \ll 1; \label{impli3} \\
& & ||S_1\Psi_0||^2 \approx ||S_1'\Psi_0||^2 \; \hbox{and} \; \sqrt{m(S_1, S_2)}, \sqrt{m(S'_1, S'_2)}, w(S_1, S'_1) \ll 1 \Rightarrow w(S_2, S'_2) \ll 1, \; \; \; \; \; \; \; \;  \label{impli4}
\end{eqnarray}
where, given two positive number $c_1$ and $c_2$, with $c_1 \approx c_2$ we mean here that, if $\epsilon \ll 1$, then $(c_1/c_2) \epsilon \ll 1$ and $ (c_2/c_1) \epsilon \ll1$ as well. We say that $c_1$ and $c_2$ are of the same order. Note that $c_1 \approx c_2$ implies $c_2/c_1, c_1/c_2 \approx 1$.

Implications (\ref{impli1}) to (\ref{impli3}) can be deduced from inequalities (\ref{ine1}) to (\ref{ine3}) respectively because $m^3(S_1, S_2) \ll 1 \Leftrightarrow m(S_1, S_2) \ll 1$ (implication (\ref{impli3}) also make use of inequality (\ref{ine1})). Implication (\ref{impli4}) derives from inequality (\ref{ine4}) due to the fact that $a, b, c \approx 1$. Indeed, assume that $\sqrt{m(S_1, S_2)}, \sqrt{m(S'_1, S'_2)} \leq \epsilon \ll 1$, and consider for instance $a$. We have:
$$
a=\frac{\max\{||S_1\Psi_0||,||S_2\Psi_0||\} }{||S_2\Psi_0||} \min \left \{\frac{||S_2\Psi_0||}{||S_1\Psi_0||}\frac{||S'_1\Psi_0||}{||S'_2\Psi_0||}\frac{||S_1\Psi_0||}{||S'_1\Psi_0||},\frac{||S_1\Psi_0||}{||S_2\Psi_0||}\frac{||S'_2\Psi_0||}{||S'_1\Psi_0||}\frac{||S'_1\Psi_0||}{||S_1\Psi_0||} \right \}.
$$

We have 
$$
1- \epsilon  \leq \frac{\max\{||S_1\Psi_0||,||S_2\Psi_0||\} }{||S_2\Psi_0||}, \frac{||S_2\Psi_0||}{||S_1\Psi_0||}, \frac{||S'_1\Psi_0||}{||S'_2\Psi_0||} \leq \frac{1}{1- \epsilon}.
$$
Thus $a$ is the product of four numbers which are of the order of unity, and therefore $a$ is also of the same order.

Implication (\ref{impli3}) guarantees that the example discussed at the beginning of this subsection is satisfied. Implication (\ref{impli4}) guarantees that, if $S_1$ and $S_1'$ are non-overlapping subtree supports, also the supports of their subtrees are non-overlapping for $t \geq t_1$. In fact this result requires an assumption of the type $\epsilon \ll 1 \Rightarrow \sqrt{\epsilon} \ll 1$, which is not completely satisfactory. This is due to the fact that inequality (\ref{ine4}) contains the square root of the typicality function. Hopefully further studies will allow us to find a better inequality.

\vspace{3mm}
It is obvious that the results discussed in this section only partially solve the problem of proving the consistency of the quantum typicality rule, for which a rigorous proof remains an open problem.

\section{Discussion and conclusion} \label{discussion}

We have seen that the explanation of the observed phenomena given by a path space representing the universe (i.e. a set of paths with a probability measure defined on it) is based on the definition of: (1) the typical configurations at a fixed time relative to a subset of the configuration space, which explain the results of statistical experiments and (2) the branches, which explain the observable structure of the trajectories, i.e. the macroscopic evolution. Both these notions can be derived by a relative typicality function.

We have also seen that the quantum formalism can provide such a typicality function, without the need of an underlying probability measure. As a consequence, in place of a path space $(M^T, \mu)$, a more economic model for the universe is the pair $(M^T, \Psi)$, where $\Psi$ represents the universal wave function, i.e. the initial wave function $\Psi_0$ plus the unitary time evolution operator $U(t)$ (the possible requirement for the trajectories to have a well defined asymptotic velocity is ignored in this section). In order to make the model more palatable, the set $M^T$ could be replaced by the set $M^T_C$ of the continuous functions, even if this replacement has no empirical consequence. 

\vspace{3mm}
It is natural to attribute to $M^T$ and $\Psi$ a meaning analogous to that of the elements $M^T$ and $\mu$ of a canonical stochastic process. The presence of $M^T$ endows the model with a definite ontology, and allows us to think that the particles of the universe follow definite trajectories, even if there are theoretical limits to our possibility to know them. These limits depend on the possibility of recording knowledge. On the other hand, the universal wave function $\Psi$ would have to be considered something like a probability measure, even if it contains less structure than a probability measure; namely, in place of the detailed finite dimensional distributions, it provides correlations between two different time s-sets in terms of mutual typicality. These correlations are expressed by the quantum typicality rule. Analogously to a canonical stochastic process, the set $M^T$ has no empirical content, i.e. any empirical prediction provided by the model can be derived from the only universal wave function. However, removing $M^T$ from the model for this reason would not be a good idea, in the same way in which removing $M^T$ from a canonical stochastic process is not a good idea. The presence of $M^T$ gives logical coherence to the model; by removing it one would obtain the Many World Interpretation, with its well known conceptual and interpretative problems.

\vspace{3mm}
The formulation of quantum mechanics proposed in this paper has the merits of Bohmian mechanics, namely the solution of the measurement problem, the explanation of the emergence of a classical world, and the presence of a non-vague ontology. On the other hand, this formulation does not make use of  the guidance equation and of the related trajectories, which, due to their non-observability, are sources of many controversies. One can say that the guidance equation is replaced by the quantum typicality rule.

\vspace{3mm}
In fact, what has been argued in this paper is that the pair  $(M^T, \Psi)$, together with the quantum typicality rule, can {\it potentially} explain the observed phenomena, but it has not been proved that it {\it actually} explains them. In order to prove this, one must prove that the model gives rise to (i) the expected results for the statistical experiments, and (ii) to a quasi-classical structure for typical trajectories. As to the first request, most of the work has already been done by proving the quantum equilibrium hypothesis \cite{durr1}. As to the second request, since typical trajectories follow the branches of the universal wave function, one must prove that the universal wave function actually has a branch structure, and that the branches have a quasi-classical structure. We have not confronted this problem in this paper.

\section{Acknowledgments}
The author wants to thank N. Zangh\`{\i} for a useful discussion and encouragement.

%end{large}

\end{document}